\documentclass[11pt,twoside]{article}

\usepackage{asp2014}

\aspSuppressVolSlug
\resetcounters

\begin{document}

\title{APEX Control System (APECS): Recent improvements and plans}

\author{Dirk Muders,$^1$ Carsten K\"onig,$^1$ Reinhold Schaaf,$^2$ Felipe Mac-Auliffe,$^3$ and Juan-Pablo P\'erez-Beaupuits$^3$}
\affil{$^1$Max-Planck-Institut f\"ur Radioastronomie, Bonn, Germany; \email{dmuders@mpifr.de}}
\affil{$^2$Argelander-Institut f\"ur Astronomie, Bonn, Germany}
\affil{$^3$European Southern Observatory, Santiago de Chile, Chile}

\paperauthor{Dirk Muders}{dmuders@mpifr.de}{0000-0002-2315-2571}{Max-Planck-Institut f\"ur Radioastronomie}{Submillimeter Technology}{Bonn}{}{53121}{Germany}
\paperauthor{Carsten K\"onig}{koenig@mpifr.de}{0000-0001-8032-9874}{Max-Planck-Institut f\"ur Radioastronomie}{Submillimeter Technology}{Bonn}{}{53121}{Germany}
\paperauthor{Reinhold Schaaf}{rschaaf@astro.uni-bonn.de}{}{Argelander-Institut f\"ur Astronomie}{}{Bonn}{}{53121}{Germany}
\paperauthor{Felipe Mac-Auliffe}{fauliffe@eso.org}{}{European Southern Observatory}{APEX}{Santiago de Chile}{}{19001}{Chile}
\paperauthor{Juan-Pablo P\'erez-Beaupuits}{jperezbe@eso.org}{0000-0003-3536-2274}{European Southern Observatory}{APEX}{Santiago de Chile}{}{19001}{Chile}



\begin{abstract}
We report on recent improvements of the Atacama Pathfinder Experiment Control System (APECS)
to cope with the ever increasing data rates and volumes. Also the very wide bandwidths of
current instruments required switching to vectorized atmospheric opacity corrections using
parallelization to speed these computations up for the quasi-realtime online pipeline. We look
ahead at the coming years of continued APEX operations.
\end{abstract}



\section{Overview}
APEX (Atacama Pathfinder EXperiment) is a 12m single dish submillimeter telescope at 5100m
altitude near the ALMA site on the Chajnantor plateau in Chile. The APEX Control System
(APECS) provides telescope control, instrument configuration, raw data acquisition and data
calibration which is handling observing sequences (e.g. on/off) and instrumental or
atmospheric corrections. It also implements automatic observation and device parameter
logging. The user interface is based on Python and allows for efficient scripted observations.

Due to the generic interfaces, APECS has been able to accommodate all facility and PI
instruments during the past 15 years. In recent time many improvements were made to handle
the ever increasing data rates and volumes. New calibration schemes can cope with the very
wide bandwidths of current and future instruments. APECS was also adapted to use the new
wobbler and hexapod of the APEX-II telescope hardware upgrade.

Future developments will address porting the software to Python 3 and migrating the
real-time computer to a modern platform.

\section{APECS Design}

APECS\footnote{APECS Manual
(https://www.mpifr-bonn.mpg.de/technology/submm/apecs\_manual\_latest)} \citep{2006A&A...454L..25M}
is distributed across a number of servers that communicate with each other using
the middleware provided by the ALMA Common Software (ACS, \citet{2013ascl.soft02003C}). ACS implements an object
oriented component model with automatic parameter monitoring, logging and alarm services,
and communication channels. ACS is based on the Common Object Request Broker Architecture
(CORBA\footnote{https://www.corba.org}). Figure \ref{fig1} shows the overall APECS design.

APECS defines generic interface classes for frontends (heterodyne \& continuum), backends
(spectrometers \& total power recorders) and auxiliary devices such as Intermediate
Frequency Processors (IFs). These interfaces allow adding new instruments very easily since
the actual observing software does not have to be modified. The common interfaces are geared
toward the astronomy rather than the technical domain.

\articlefigure{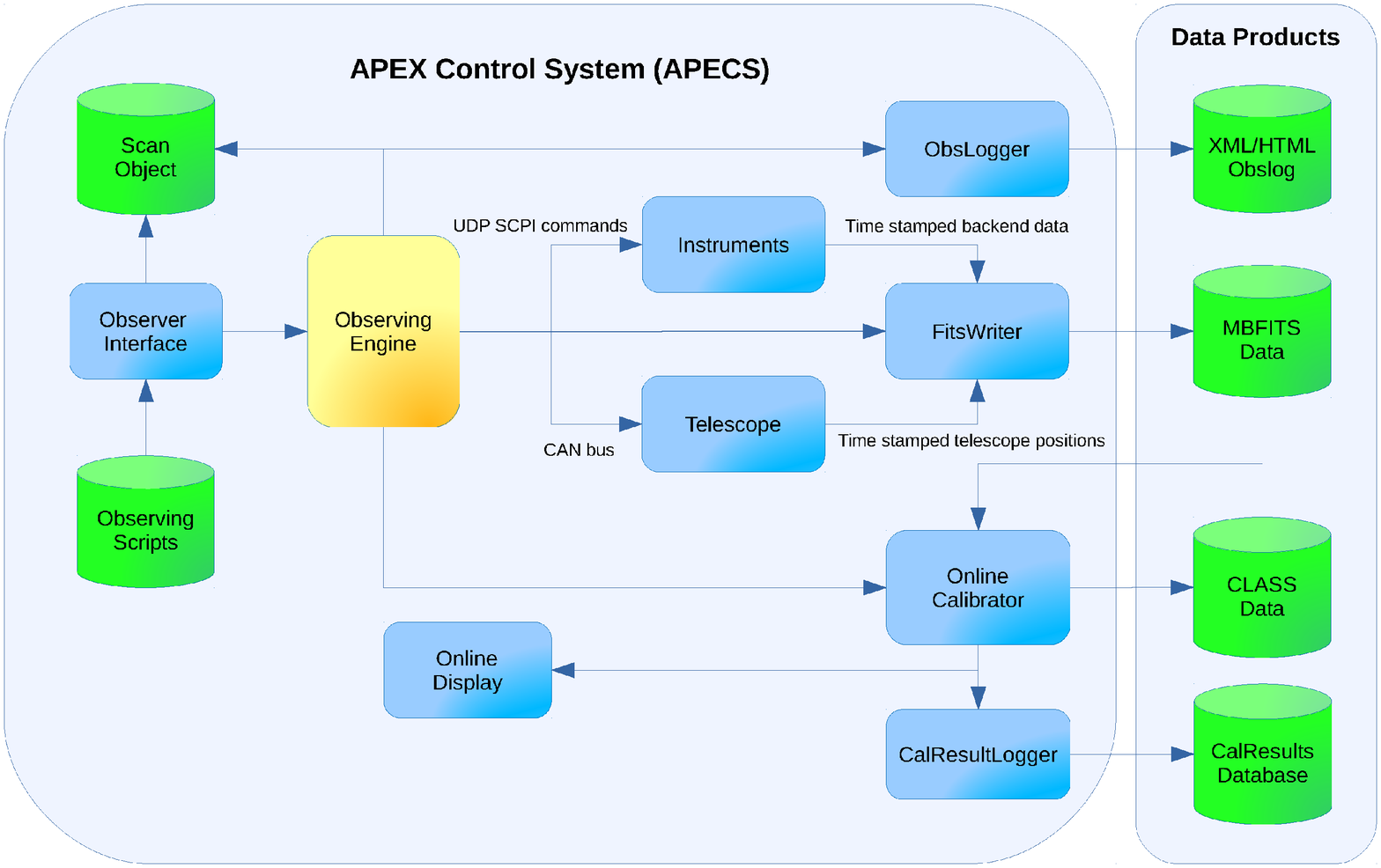}{fig1}{APECS design with major components and connections.}

Communication to the hardware controllers uses text commands according to the UDP SCPI
protocol. This facilitates mapping the APECS object oriented interfaces to even very small
micro controller based systems.

In APECS the "Observing Engine" uses a scan description object to configure telescope and
instruments as required and to start and stop data taking and calibration. The observer sets
up the scans using a Python shell with special commands. A remote control mode allows
interfacing APECS to the VLBI field system e.g.\ for EHT observations.

\section{Large Data Handling}

The amount of data produced by APEX instruments has increased exponentially over time.
In 2019 the maximum data rate was of the order 35 MB/s and a total of about 70 TB of raw
and calibrated data were accumulated. The next generation of frontends and backends will
push these limits further up.

The original APECS setup was based on 32 bit operating systems which impose a memory limit
of 3 GB per process. The old numerical library used in Python ("Numeric") had an even lower
limit of 2 GB per array variable. As a consequence, some observing modes like fast On-The-Fly
mapping had to be restricted in size and duration to be able to calibrate the data. The
switch to 64 bit systems was thus urgently needed.

The porting affected all software modules and took place over a period of more than a year.
In addition to general 32 to 64 bit conversions, the numerical library had to be updated to
"numpy" in the most important applications like the Calibrator and part of the Multi-Beam FITS
(MBFITS\footnote{MBFITS Format Definition
(https://www.mpifr-bonn.mpg.de/technology/submm/mbfits\_latest)}) Writer.
The new system APECS 4.0 was deployed during the 2018/19 shutdown time. Apart from removing
the memory limitations, the switch to
"numpy" also provided a significant speed-up of the online calibration. In 2020 the remainder
of applications was ported to "numpy" in APECS 4.1.

\section{Parallelized Data Writing}

The MBFITS Writer used to be a multi-threaded Python application. The Python threading model
is limited by the so called "Global Interpreter Lock" (GIL) which couples the threads to some
extent. For data rates beyond 20 MB/s the MBFITS Writer reached the maximum load per process
and started losing data. We therefore needed to develop a multi-process parallelized
version. It decouples the TCP data receiver into a separate process. We tested this setup with
a 10 Gbit network and were able to handle spectral backend data rates up to about 300 MB/s.

\articlefigure{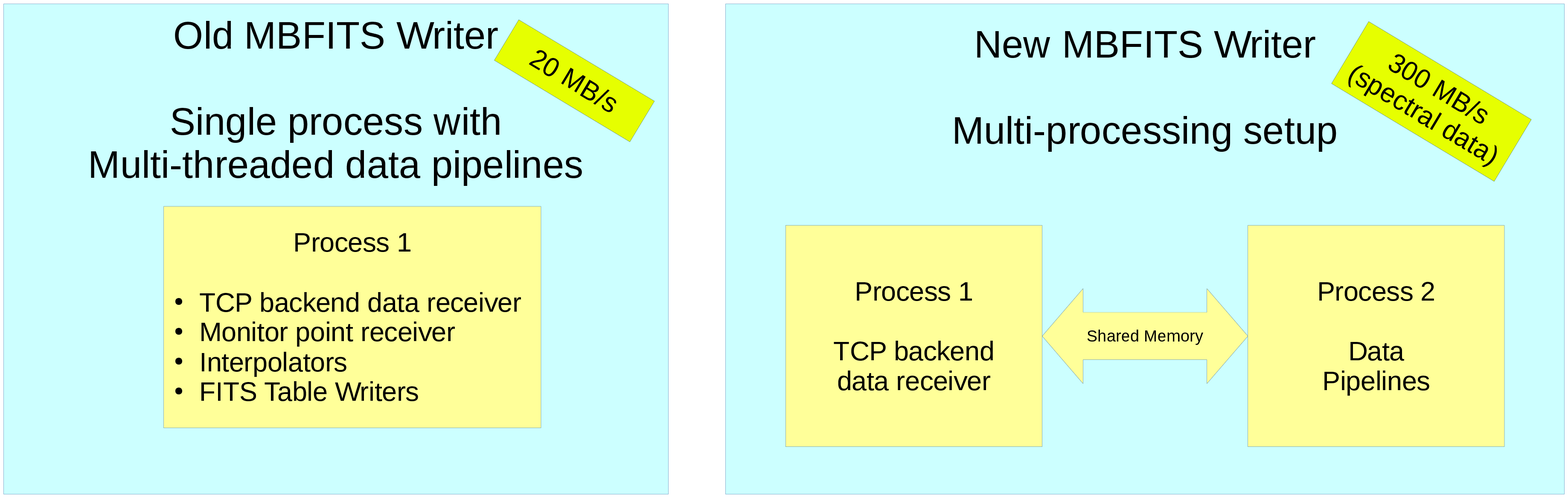}{fig2}{MBFITS Writer design and performance improvements.}

\section{Vectorized Opacities Calibration}

With ever increasing frontend and backend bandwidths we have reached the
point were the simple scalar opacity calibration is no longer adequate. In
order to compute vectorized ATM solutions, the APEX
Calibrator\footnote{APEX Calibrator Manual
(https://www.mpifr-bonn.mpg.de/technology/submm/calibrator\_manual\_latest)}
code had to be parallelized
using the Python "multi-processing" library. The online calibration now
calculates one opacity per MHz in the spectrum. The vector opacities (signal
and image band) are also written to the CLASS file.

Recently another limitation in calibrating data of the B cabin receivers
became obvious. The calibration unit provides a corrected T$_{\rm cold}$ value that
varies sinusoidally with frequency. Up until 2019 the Calibrator assumed a
single T$_{\rm cold}$ value per 4 GHz spectrum. This led to jumps between the basebands
in T$_{\rm RX}$, T$_{\rm cal}$ and the spectra. The new Calibrator of APECS 4.1, which was
deployed in January 2020, now interpolates the T$_{\rm cold}$ values across a baseband
and provides much improved calibrations (see figure \ref{fig3}).

\articlefigure{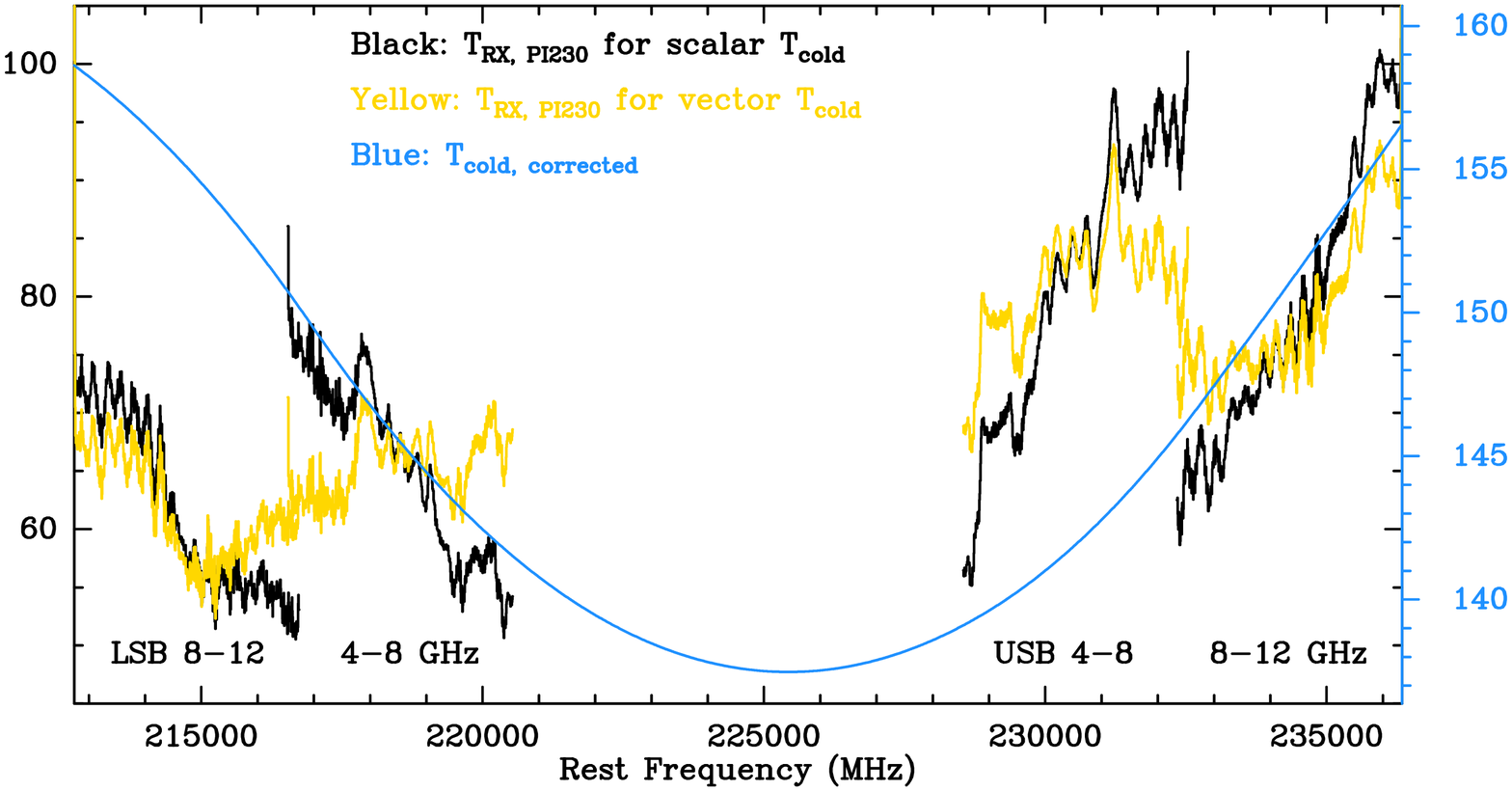}{fig3}{Vectorized T$_{\rm cold}$ calibration
avoids jumps between the IF basebands.}

\section{Future Developments}

APECS is now more than 15 years old. While ACS and Linux updates allowed
running the system on modern hardware, some of the technologies and software
packages like Python 2 are now aging significantly. In order to be able to
operate APEX for the coming years, we will work on upgrading these items.
For next generation instruments APECS will need to be
further enhanced to provide the proper data handling and calibration. This
will also involve improving the MBFITS raw data format to speed up data access
for the monitor tables.

\bibliography{P11-180}  

\end{document}